\documentclass[11pt]{article}

\usepackage[final]{acl}

\usepackage{times}
\usepackage{latexsym}
\usepackage{amsmath}

\usepackage[T1]{fontenc}
\usepackage[utf8]{inputenc}

\usepackage{microtype}

\usepackage{inconsolata}

\usepackage{graphicx}

\title{From Standard English to Singlish: A Retrieval-Augmented Approach for Code-Switched Creole Generation in Large Language Models}

\author{Foong Ming Lai, Yujin Tan, Han Meng, Yi-Chieh Lee \\
  Department of Computer Science, National University of Singapore \\
  \texttt{\{foongming, tan.yugin, han.meng\}@u.nus.edu, yclee@nus.edu.sg}}

\begin{document}
\maketitle
\begin{abstract}

Code-switching in contact varieties like Singaporean English (Singlish) challenges natural language generation due to limited parallel data and rapid lexical evolution. We propose a retrieval-augmented generation (RAG) framework that externalizes dialectal knowledge into a curated lexicon, enabling controlled lexical code-switching without fine-tuning. Our approach retrieves candidate Singlish expressions and guides generation through sparse lexical substitution. Human evaluation with 164 Singaporean participants found RAG and zero-shot prompting equally natural and appropriate. Automatic analyses reveal different transformation regimes: zero-shot prompting induces extensive paraphrasing (median 23 token edits), whereas RAG performs minimal substitutions (median 1 edit) with higher semantic preservation (mean cosine similarity 0.978 vs. 0.926). Our results demonstrate that externalizing code-switching into lexical resources enables control and auditability without sacrificing perceived quality, offering practical advantages for rapidly evolving contact varieties.

\end{abstract}

\section{Introduction and Related Work}
\label{sec:intro}

Code-switching (CS) is a multilingual phenomenon in which speakers alternate languages within the same utterance. Beyond being a purely lexical mixture, CS functions as a communicative and interactional resource, expressing socially meaningful cues \citep{dogruoz-etal-2021-survey}. Generating code-switched text is practically relevant for conversational agents, where aligning to local norms improves perceived naturalness and user experience. \citep{bawa-etal-2020-prefer}.

Most existing approaches to code-switched text generation are not well matched to low-resource language varieties such as Singaporean English (Singlish). As a contact-influenced variety \citep{wang-etal-2017-universal}, Singlish is characterized by rapid lexical innovation and sociolinguistic change \citep{hafiz-etal-2024-sia}, where locally meaningful cues are expressed through lexical choices and short stretches of mixing within otherwise English-dominant utterances. A common paradigm for CS generation uses parallel data mapping of monolingual inputs to code-switched outputs \citep{winata-etal-2019-code}, but the corpora required for this technique are unavailable for Singlish and many other contact varieties. Other approaches rely on task- or language-specific fine-tuning to induce code-switching behavior \citep{gupta-etal-2020-semi,tarunesh-etal-2021-machine}. Although small, high-quality code-switched datasets can substantially improve supervised adaptation \citep{olaleye-etal-2025-afrocs}, multilingual LLMs often remain unreliable on code-switched inputs unless explicitly adapted, with deficits reported across tasks such as sentiment analysis, translation, identification, and summarization \citep{khanuja-etal-2020-gluecos,zhang-etal-2023-multilingual}.

Fine-tuning and prompting also raise governance concerns for rapidly changing varieties like Singlish. Parameter adaptation requires retraining to track lexical drift and resists auditing at the lexical level. Unconstrained dialect generation also risks producing overly stereotypical or otherwise inappropriate content within non-standard English varieties  \citep{fleisig-etal-2024-linguistic,bui2025largelanguagemodelsdiscriminate}. These issues motivate approaches that provide explicit control over lexical choices without repeatedly modifying model parameters.

Retrieval-augmented generation (RAG) is a standard technique for grounding language models in external resources \citep{lewis2021retrievalaugmentedgenerationknowledgeintensivenlp} and has been extended to multilingual settings \citep{chirkova2024retrievalaugmentedgenerationmultilingualsettings,kruk2025banglassistbengalienglishgenerativeai}. However, this line of work has primarily focused on factual grounding and multilingual query handling rather than controlled lexical variation. We reposition retrieval as a mechanism for lexical governance in dialect generation, proposing a retrieval-augmented framework that treats code-switching as lexical control rather than parametric adaptation. Our key design choice is to externalize dialectal knowledge into an explicit, editable lexicon. At inference time, the system retrieves candidate dialectal items and conditions generation on them, enabling minimal substitutions (median 1 token edit) without parallel training data or variety-specific fine-tuning. Because lexical resources are externalized, vocabulary choices can be inspected, updated, restricted, or filtered without retraining the underlying language model.

In this work, we compare retrieval-guided lexical rewriting against zero-shot prompting using both intrinsic analyses of the induced transformation regime (edit minimality and semantic faithfulness) and human judgments of perceived Singlish use and appropriateness. We make three contributions to the NLP community. 1) We demonstrate that sparse lexical substitution and extensive paraphrasing achieve comparable user-perceived naturalness despite operating under distinct transformation regimes. 2) We show that lexical control enables sparse, localized edits with higher semantic preservation, in contrast to the extensive paraphrasing induced by zero-shot prompting. 3) We provide evidence that externalizing code-switching into lexical resources enables control and auditability without sacrificing perceived quality.

\section{Methodology}
\label{sec:method}

\begin{figure}
    \centering
    \includegraphics[width=1\linewidth]{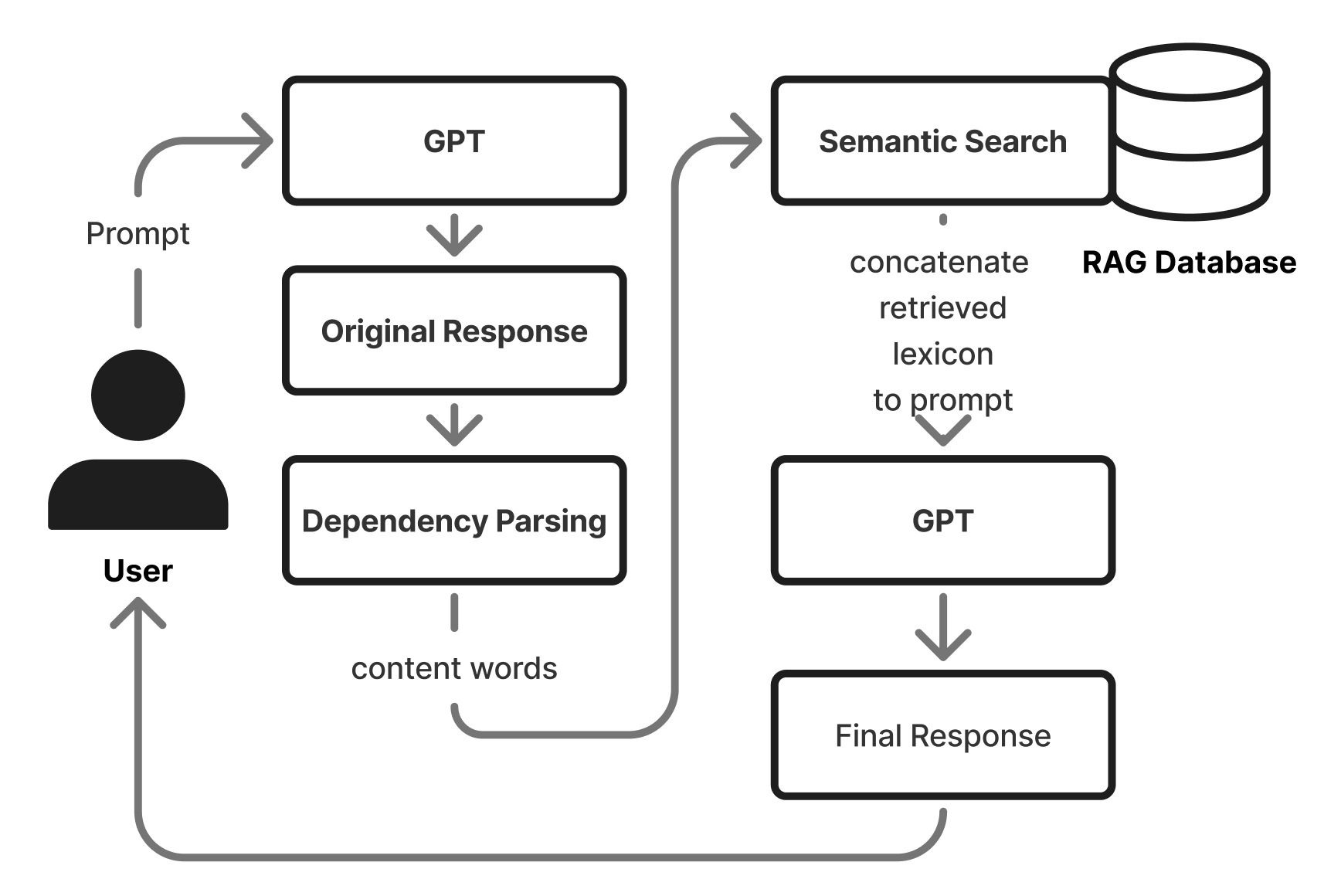}
    \caption{Overview of our retrieval-augmented lexical code-switching pipeline. Given a user prompt, a base GPT model first produces an English response; we dependency-parse this response to extract content words, use them to query a Singlish lexicon via semantic search, then concatenate the retrieved lexical entries back into the prompt for a second GPT pass that generates the final, lexically code-switched response.}
    \label{fig:pipeline}
\end{figure}

We propose a retrieval-augmented pipeline illustrated Fig. \ref{fig:pipeline} that externalizes code-switching into a curated lexicon.

Given a dialogue context $C$ and user utterance $U$, we generate a response $Y$ that is semantically coherent and incorporates lexical code-switching into Singlish. We assume no parallel data (Standard English$\rightarrow$Singlish) and no variety-specific fine-tuning; the only resource is a discrete lexicon $L$ of Singlish forms with glosses and usage examples. We focus on \textbf{lexical} code-switching - sparse insertion/substitution of content words and fixed expressions (e.g., \textit{sian}, \textit{paiseh}) -and exclude broader syntactic restructuring. The goal is to test whether retrieval alone can support controllable, localized dialectal edits rather than unconstrained paraphrasing.

{An example of code switching is as follows:}
\begin{quote}
    Original: ``Hi there, that sounds really \textbf{exhausting}. Back-to-back meetings can wear anyone down, and it's understandable to feel overwhelmed.''
\end{quote}
\begin{quote}
    Rewritten: ``Hi there, that sounds really \textbf{sian}. Back-to-back meetings can wear anyone down, and it's understandable to feel overwhelmed.''
\end{quote}

% \section{Method: Lexical Retrieval-Augmented Code-Switching}

\subsection{Lexicon Construction}

The method assumes access to a lexicon $L$ of expressions characteristic of a target code-switched variety $V$. Each entry in $L$ consists of (i) a code-switched term, (ii) a Standard English gloss, and (iii) a usage example. In our experiments, $L$ contains 198 common Singlish expressions, taken from a popular Singlish dictionary\footnote{https://singlishdict.app/}. The lexicon is treated as a standalone resource and can be modified without retraining the language model.

All lexicon entries are embedded using OpenAI \texttt{text-embedding-3-small} and indexed in a shared embedding space using an approximate nearest-neighbor (ANN) HNSW index implemented in \texttt{hnswlib} with cosine similarity.

\subsection{Generation and Candidate Retrieval}

Given a dialogue context $C$, a base large language model (LLM) first generates an initial response $r^{\text{EN}}$ in Standard English using a fixed system prompt.

To identify candidate substitution sites, we extract content words from $r^{\text{EN}}$ using dependency parsing, retaining tokens with POS tags corresponding to NOUN, PROPN, VERB, ADJ, ADV. 
% , proper nouns, verbs, adjectives, and adverbs. These tokens serve as semantic anchors for lexical retrieval.

Each extracted content word is embedded using the same embedding model (\texttt{text-embedding-3-small}) and queried against the lexicon index via ANN. 

% For each query, up to three candidate code-switched expressions are retrieved. To maximize recall under low-resource conditions, we apply a low similarity threshold and over-generate candidates, which are then deduplicated, ranked by cosine similarity, and truncated to a maximum of ten unique expressions.

\subsection{Rewrite with Lexical Cues}

The retrieved code-switched expressions and their English glosses are appended to the input as lexical context. The model is prompted to rewrite the initial response $r^{\text{EN}}$ in the target variety \textit{V}, with the retrieved items serving as candidate substitutions.

\section{Experimental Setup}

{\it Conditions.}
We compare three generation conditions that differ only in how code-switching is induced:
\textbf{a) Baseline (Standard English):} The model generates responses in Standard English without any instruction to code-switch.
\textbf{b) RAG (Lexical):} Our proposed method, where code-switching is guided by retrieval from the Singlish lexicon $L$ followed by a rewrite step.
\textbf{c) Zero-Shot (Prompting):} The model is instructed via prompting alone to respond in natural Singaporean English, relying on its internal parametric knowledge.

All conditions use the same base model (GPT-5), system prompt, and decoding parameters. This isolates the effect of the code-switching control mechanism from other model or interface factors.

% \subsection{Research Design}

% We conducted a between-subjects user study ($N=164$) to evaluate the proposed method against standard prompting baselines. Participants were randomly assigned to one of three conditions, differing only in the \textbf{code-switching control mechanism}:\begin{itemize}\item \textbf{Baseline (Standard English):} The model generates responses in Standard English with no instruction to code-switch.\item \textbf{Zero-Shot (Holistic):} The model is prompted to reply in "natural Singaporean English" using its internal parametric knowledge, without external retrieval.\item \textbf{RAG (Lexical):} Our proposed method, where code-switching is induced via retrieval from the lexicon $L$ and a subsequent rewrite step.\end{itemize}All conditions utilized the same base model (GPT-5), system persona (empathetic support), and decoding parameters. This design isolates the impact of the control mechanism—internal knowledge vs.\ external retrieval—on perceived naturalness and appropriateness.

%\subsection{Participants}

{\it Evaluation.} 164 Singaporeans took part in the study. Participants who did not complete the interaction or failed an embedded attention check were excluded prior to analysis.

System outputs were evaluated through human judgments collected after the interaction. Participants rated the chatbot on 7-point Likert scales measuring:
1) Perceived Singlish Use (USE): the extent to which the chatbot used Singlish;
2) Appropriateness (APPR): whether the Singlish usage was perceived as correct and natural.

\section{Results}

\subsection{Edit Minimality}

\paragraph{Edit distance measures transformation scope.}
Because our method is designed as \emph{lexical code-switching without syntactic restructuring}, the primary intrinsic question is not ``does the output match a reference,'' but ``how \emph{invasive} is the rewrite.'' We therefore quantify modification magnitude using token-level Levenshtein distance, i.e., the minimum number of token insertions, deletions, and substitutions required to transform the original response into the generated response \citep{levenshtein1966binary}. This aligns with work treating generation as text editing. \citep{malmi-etal-2019-encode,mallinson-etal-2020-felix}. In our experiments we lowercase tokens while retaining punctuation to avoid capitalization artifacts while still reflecting structural changes (e.g., clause insertion or reordering) in the distance.

% To quantify how invasive each transformation is, we compute word-level Levenshtein edit distance between the original English response and the generated output. Tokens are lowercased while punctuation is retained, ensuring that capitalization artifacts are ignored while sentence-level structural changes are preserved.

Table~\ref{tab:edit_distance} reports edit-distance statistics for zero-shot prompting and the proposed RAG method. Zero-shot prompting induces extensive rewriting, with a median of 23 token edits and virtually no unchanged outputs. In contrast, the RAG method performs sparse, localized edits: the median edit distance is 1 token, with over 92\% of outputs differing by at most two tokens.

These results indicate that zero-shot prompting operates primarily as a global paraphrasing mechanism, whereas the proposed RAG approach functions as a controlled lexical rewrite operator, intervening only when suitable dialectal substitutions are available.

\begin{table*}[t]
\small
  \centering
  \begin{tabular}{lrrrrr}
    \hline
    \textbf{Method} & \textbf{$N$} & \textbf{Median edits} & \textbf{Mean edits} & \textbf{\% $\leq$2 edits} & \textbf{\% $\leq$5 edits} \\
    \hline
    Zero-shot & 1266 & 23 & 24.7 & 0.1 & 1.3 \\
    RAG & 838 & 1 & 1.23 & 92.6 & 99.6 \\
    \hline
  \end{tabular}
  \caption{\label{tab:edit_distance}
    Token-level edit distance between original and generated responses (lowercased word tokens; punctuation retained).
    Zero-shot prompting induces extensive rewriting, whereas RAG performs sparse, localized lexical edits.
  }
\end{table*}

\subsection{Semantic Faithfulness}

% \paragraph{Why embedding similarity is a reasonable proxy for semantic faithfulness.}
% In controlled rewriting problems such as text style transfer, evaluation is commonly decomposed into (i) strength of the targeted change, (ii) meaning/content preservation, and (iii) fluency. Because reference outputs are often unavailable, a standard practice is to approximate meaning preservation via embedding-based similarity between the source and the transformed output \citep{fu2017styletransfertextexploration,briakou-etal-2021-evaluating}. 

Minimal edits are only desirable insofar as they preserve meaning. For our case, because reference outputs are unavailable, a standard practice is to approximate meaning preservation via embedding-based similarity between the source and the transformed output \citep{fu2017styletransfertextexploration,briakou-etal-2021-evaluating}. To assess semantic faithfulness, we compute cosine similarity between sentence embeddings of the original and generated responses using the \texttt{text-embedding-3-small} model.

As shown in Table~\ref{tab:semantic_similarity}, RAG outputs remain extremely close to the original responses in embedding space (mean cosine similarity 0.978; median 0.991). Zero-shot prompting exhibits lower semantic similarity and a heavier lower tail (5th percentile 0.849), reflecting occasional semantic drift associated with extensive paraphrasing.

Across both methods, edit distance and semantic similarity are negatively correlated ($r=-0.20$ for zero-shot; $r=-0.34$ for RAG), indicating that tighter control over the extent of rewriting is associated with improved semantic faithfulness.

\begin{table*}[t]
\small
  \centering
  \begin{tabular}{lrrrr}
    \hline
    \textbf{Method} & \textbf{$N$} & \textbf{Mean cosine} & \textbf{Median cosine} & \textbf{5th percentile} \\
    \hline
    Zero-shot & 1266 & 0.926 & 0.937 & 0.849 \\
    RAG & 838 & 0.978 & 0.991 & 0.915 \\
    \hline
  \end{tabular}
  \caption{\label{tab:semantic_similarity}
    Semantic faithfulness measured as cosine similarity between sentence embeddings of original and generated responses.
    RAG preserves meaning more faithfully than zero-shot prompting.
  }
\end{table*}

\subsection{Edit Distance vs.\ Semantic Similarity}

\begin{figure}
    \centering
    \includegraphics[width=1\linewidth]{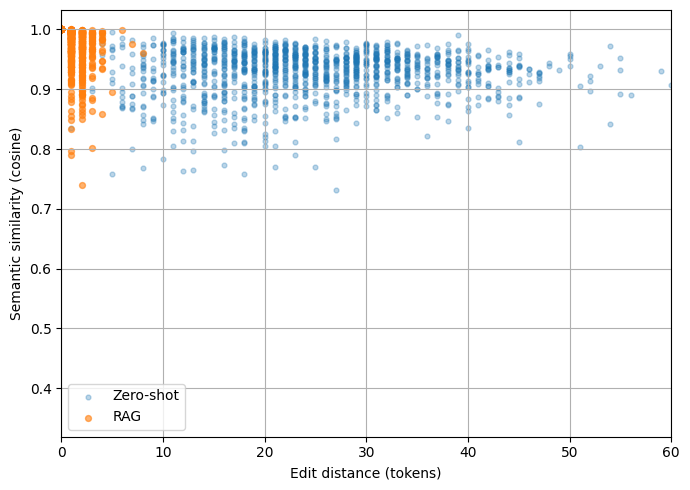}
    \caption{Edit distance versus semantic similarity for zero-shot prompting and RAG. The x-axis is capped for readability.}
    \label{fig:edit_vs_semantic}
\end{figure}

Figure~\ref{fig:edit_vs_semantic} illustrates the relationship between edit distance and semantic similarity. Zero-shot prompting exhibits large edits and greater semantic variability, whereas RAG outputs cluster tightly around small edit distances with high semantic similarity. This pattern shows that the two approaches differ not only in outcome but in the \emph{type of transformation} they perform.

\subsection{Human Evaluation}
\label{sec:human_eval}

This study received IRB approval. Participants (N=164) were recruited via Telegram, provided informed consent, and received SGD 6 compensation. All Telegram identifiers were replaced with system-generated anonymous IDs during data collection, and analyses were conducted solely on anonymized conversation logs. Both RAG and Zero-shot received substantially higher ratings than Baseline on both dimensions of perceived use and appropriateness of use. Independent samples $t$-tests confirmed significantly higher USE ratings for RAG ($p < .001$, $d = 2.48$) and Zero-shot ($p < .001$, $d = 2.67$) compared to Baseline. Similarly, APPR ratings were significantly higher for RAG ($p = .002$, $d = 0.64$) and Zero-shot ($p < .001$, $d = 1.09$) than Baseline. No significant differences were observed between RAG and Zero-shot on either USE ($t(113) = -0.21$, $p = .83$) or APPR ($t(112) = -1.52$, $p = .13$).

These results indicate that retrieval-based and prompting-based approaches achieve comparable user-perceived levels of Singlish use and appropriateness, despite differing substantially in their underlying transformation behavior.

\section{Discussion}

Our results show both RAG and zero-shot prompting successfully induce perceptible and appropriate Singlish, with no significant differences in aggregate user judgments. However, automatic analyses reveal that the two approaches operate under fundamentally different transformation regimes. Zero-shot prompting relies on unconstrained paraphrasing entangled with prompt instructions and model internals, whereas RAG externalizes code-switching into an explicit lexical resource, enabling sparse, localized edits and greater control over vocabulary use. This design offers practical advantages for rapidly evolving varieties like Singlish. While RAG does not outperform prompting on perceptual measures, it provides a more modular and maintainable mechanism for inducing code-switching without modifying model parameters.
\section{Future Work}

A natural next step is to apply the proposed framework to other English-lexifier creoles and contact varieties (e.g., Malaysian English (Manglish)) by constructing corresponding lexicons and evaluating retrieval-guided code-switched generation in those settings. 

%As the approach decouples code-switching behaviour from model parameters, extending it to new varieties primarily involves curating lexical resources rather than retraining models.

% A complementary direction is to invert the current setup and inject English into conversations that are primarily conducted in another language (e.g., Mandarin--English or Korean--English code-switching). This would require integrating non-English parsers and lexicons into the retrieval pipeline and adapting the rewrite step to non-English matrix languages.

% Finally, future work should incorporate expert or native-speaker evaluations to assess finer-grained aspects of code-switched output, including grammaticality, pragmatic appropriateness, and sociolinguistic naturalness. Such evaluations would complement user-based judgments and help clarify where retrieval-guided lexical control succeeds or fails relative to prompting-based approaches.

\newpage
\section{Limitations}

This study focuses on lexical borrowing to isolate semantic appropriateness, explicitly excluding discourse particles and syntactic phenomena. As a result, the current framework does not capture the full range of grammatical or pragmatic features characteristic of Singlish.

Our implementation also assumes an English-centric pipeline, relying on English parsing tools and treating English as the matrix language. While this is appropriate for English-based mixed varieties such as Singlish, it remains unclear how well the approach would generalize to settings where the matrix language is non-English.

\section{Ethical Considerations}
While our lexicon was carefully curated to exclude derogatory and offensive expressions, the externalized design that enables beneficial updates also introduces risk: implementers could modify the lexicon to include harmful terms, enabling the generation of inappropriate code-switched text.

% Finally, our evaluation relies on user judgments rather than expert linguistic annotation. While these judgments capture perceived recognizability and correctness, they may obscure finer-grained distinctions in grammaticality or sociolinguistic nuance.

%\section*{Ethics Statement}
%Our work focuses on controlled generation, which mitigates the risk of LLMs generating offensive dialectal stereotypes. However, we acknowledge that reducing a dialect to a lookup lexicon is a simplification that may not capture the full cultural depth of the language.

\section*{Acknowledgments}

We are grateful to the creators and maintainers of the Singlish Dictionary (\texttt{singlishdict.app}) for granting us permission to use their data in this research. We also thank the study participants for their time and insights. This work was partially supported by DSO National Laboratories.

% Bibliography entries for the entire Anthology, followed by custom entries
%\bibliography{anthology,custom}
% Custom bibliography entries only
\bibliography{custom}

\appendix

\section{Prompt Templates}
\label{sec:appendix}
Full prompt used when rewriting an utterance into Singlish: 

You are a linguistics expert specialising in the nuances of Singaporean English. Rewrite the message in \texttt{<TARGET>} in everyday Singaporean English. The messages are currently from a western society and are slightly awkward to hear in Singapore. Most Singaporeans speak fluent, standard English, with light Singlish touches here and there. Not broken grammar everywhere. A dictionary is provided to give you possibilities for words to replace with Singlish variants.  

Rules:
\begin{itemize}
\item You need not use any of the words from the dictionary, but you may use one of them if it fits the context.
\item Absolutely never use any discourse particles such as but not limited to la, lor, leh.
\item Replace at most ONE word with a Singlish variant for each sentence even if multiple dictionary matches are present.
\item Keep all other parts of the message exactly the same.
\item If no suitable match is found, return the original message unchanged.
\item You must preserve the message's meaning.
\item Return ONLY the rewritten message --- do not explain or add anything else.
\end{itemize}

Examples:

Dictionary: \texttt{[{{'token': 'behavior', 'word': 'pattern', 'label': 'noun','meaning': 'troublesome or annoying actions'}}, {{'token': 'driving', 'word': 'chiong', 'meaning': 'to charge; to rush forward; to make a dash for'}}]}

\texttt{<TARGET>}His behavior is driving everyone mad.\texttt{</TARGET>}\\
Rewritten: His pattern is driving everyone mad.

Dictionary: \texttt{[{{'token': 'mind', 'word': 'eye power', 'label': 'noun','meaning': 'just watching and not helping'}}]}

\texttt{<TARGET>}If there's anything on your mind, just say it.\texttt{</TARGET>}\\
Rewritten: If there's anything on your mind, just say it.

Dictionary: \texttt{[{{'token': 'must', 'word': 'die die', 'label': 'adverb', 'meaning': 'absolutely; no matter what; even at the cost of one's life'}},{{'token': 'amazing','word': 'power','meaning': 'used to express amazement, praise, etc. at something impressive or outstanding'}}]}

\texttt{<TARGET>}You must try this dish, it's amazing!\texttt{</TARGET>}\\
Rewritten: You die die must try this dish, it's amazing!

Dictionary: \texttt{[{{'token': 'act blur', 'word': 'pretend', 'label': 'verb', 'meaning': 'to feign ignorance; to play dumb'}}, {{'token': 'pretend', 'word': 'act chio', 'meaning': 'to behave in an (often exaggeratedly) charming or vain manner; to act pretty; to pretend as if one is extremely beautiful'}}]}

\texttt{<TARGET>} Don't pretend you don't know anything about the situation.\texttt{</TARGET>}\\
Rewritten: Don't act blur that you don't know anything about the situation.

Please complete the following:

Dictionary: \texttt{{dict\_str}}\\
\texttt{<TARGET>}{sentence}\texttt{</TARGET>}\\
Rewritten:

\end{document}